%% file: main_arXiv.tex
\documentclass[onecolumn]{pasj01}
\begin{document} 
\Received{2017/06/26}%{yyyy/mm/dd}
\Accepted{2017/11/13}%{yyyy/mm/dd}
%\Published{yyyy/mm/dd}

\input{definitions}

\input{s0_title}
\maketitle
\input{s0_abstract}
% * <mikito@a.phys.nagoya-u.ac.jp> 2017-06-13T14:48:47.866Z:
%
% ^.
% * <mikito@a.phys.nagoya-u.ac.jp> 2017-06-13T14:48:45.989Z:
%
% ^.
\input{s1_introduction_arxiv}
\input{s2_observations}

\input{s3_results_arxiv}
\input{s4_discussion_arxiv}

\input{s5_conclusions}
\input{s9_references_arxiv}
\input{s9_figures_arxiv}
\input{s9_tables}

\end{document}

%% file: definitions.tex
\def\Herschel{{\it Herschel}}
\def\Spitzer{{\it Spitzer}}
\def\WISE{{\it WISE}}
\def\AKARI{{\it AKARI}}

\def\um{$\mu \rm{m}$}

\def\cmcm{$\rm{cm}^{-2}$}
\def\cmcmcm{$\rm{cm}^{-3}$}
\def\kms{$\rm{km}$ $\rm{s}^{-1}$}
\def\vlsr{$v_{\rm LSR}$}
\def\degree{$^{\circ}$}
\def\Lsun{$L_{\solar}$}
\def\Msun{$M_{\solar}$}
\def\Msunyr{$M_{\solar}$ $\rm{yr}^{-1}$ }

\def\NH{$N(\rm{H_{2}})$}

\def\HII{H \emissiontype{II}}

\def\lbeq{($l$, $b$)$=$}
\def\lbsim{($l$, $b$)$\sim$}

\def\COa{\atom{C}{}{12}\atom{O}{}{}}
\def\COb{\atom{C}{}{13}\atom{O}{}{}}
\def\COc{\atom{C}{}{}\atom{O}{}{18}}

\def\Ja{{\it J}$=1$--$0$}
\def\Jb{{\it J}$=2$--$1$}
\def\Jc{{\it J}$=3$--$2$}

%% file: s0_title.tex
\title{Formation of the young compact cluster GM 24 triggered by a cloud-cloud collision }

%%% begin:list of authors
% Do NOT capitalize all letters in "textsc".
\author{Yasuo \textsc{Fukui}\altaffilmark{1,2,*}}%
\author{Mikito \textsc{Kohno}\altaffilmark{1*}}
\author{Keiko \textsc{Yokoyama}\altaffilmark{1}}%
\author{Atsushi \textsc{Nishimura}\altaffilmark{1}}%
\author{Kazufumi \textsc{Torii}\altaffilmark{3}}%
\author{Yusuke \textsc{Hatorri}\altaffilmark{1}}%
\author{Hidetoshi \textsc{Sano}\altaffilmark{1,2}}%
\author{Akio \textsc{Ohama}\altaffilmark{1}}%
\author{Hiroaki \textsc{Yamamoto}\altaffilmark{1}}%
\author{Kengo \textsc{Tachihara}\altaffilmark{1}}%
%%% end:list of authors

\altaffiltext{1}{Department of Physics, Nagoya University, Furo-cho, Chikusa-ku, Nagoya, Aichi 464-8601, Japan}
\altaffiltext{2}{Institute for Advanced Research, Nagoya University, Furo-cho, Chikusa-ku, Nagoya 464-8601, Japan}
\altaffiltext{3}{Nobeyama Radio Observatory, National Astronomical Observatory of Japan (NAOJ), National Institutes of Natural Sciences (NINS), 462-2, Nobeyama, Minamimaki, Minamisaku, Nagano 384-1305, Japan}

\email{fukui@a.phys.nagoya-u.ac.jp}
\email{mikito@a.phys.nagoya-u.ac.jp}

%% `\KeyWords{}' always has to be placed before `\maketitle'.
 %Do NOT move this preamble from here!

%% file: s0_abstract.tex
\begin{abstract}
High-mass star formation is an important step which controls galactic evolution.
GM 24 is a heavily obscured star cluster including a single O9 star with more than $\sim$100 lower mass stars within a 0.3 pc radius toward $(l,b)\sim$ (\timeform{350.5D}, \timeform{0.96D}), close to the Galactic min-starburst NGC 6334.
We found two velocity components associated with the cluster by new observations of \COa \Jb \ emission, whereas the cloud was previously considered to be single. 
We found the distribution of the two components of $5$ \kms \ separation shows complementary distribution; the two fit well with each other if a relative displacement of 3 pc is applied along the Galactic plane. 
A position-velocity diagram of the GM 24 cloud is explained by a model based on the numerical simulations of two colliding clouds, where an intermediate velocity component created by collision is taken into account.
We estimate the collision time scale to be $\sim$Myr in projection of a relative motion titled to the line of sight by 45 degrees.
The results lend further support for cloud-cloud collision as an important mechanism of high-mass star formation in the {Carina-}Sagittarius Arm.
\end{abstract}

%% file: s1_introduction_arxiv.tex
\section{Introduction}
{GM 1-24 (hereafter GM 24) is a cometary nebula cataloged by the Palomar Sky Survey (	
Gyul$'$Budagyan \& Magakian 1977)}, with an associated compact star cluster and is located at \lbsim (\timeform{350.5D}, \timeform{0.96D}) close to the min-starburst {NGC 6334} in the Carina-Sagittarius Arm (see for a review \citet{tap08}).
The brightest star is Irs3, an O9 or B0 star, and Irs3 and another nearby star Irs27 are mainly ionizing an extended \HII \ region RCW126 \citep{rod60}.
The cluster has a compact size of 0.3 pc radius, consisting of at least $\sim$100 stars, indicating high stellar density of 1000 pc$^{-3}$ with a heavy extinction A$_{\rm v} \sim$ 50--90 mag (Tapia \& Persi 2008). 
\citet{tap85} made CO \Ja \ observations and covered a molecular cloud at a resolution of \timeform{3'} with an under-sampling for an area of \timeform{20'}$ \times $ \timeform{40'} in R.A. and Dec. 
High resolution molecular observations of a small area of \timeform{200"} $\times$ \timeform{200"} were made in \COa, \COb, and \COc \ \Jb \ emissions at \timeform{22"}--  \timeform{46"} resolution with {the Swedish-ESO Submillimeter Telescope} (SEST) \citep{gom90}. 
The molecular cloud has a size of 7 pc $\times$ 14 pc , which is singly peaked at $\sim 10$ \kms, and a kinematic distance is estimated to be 2.0 kpc (Torrelles et al. 1983, G\'omez et al. 1990).

In our previous paper, CO observations were presented for the min-starbursts {NGC 6334} and {NGC 6357} next to GM 24 in the {Carina-}Sagittarius Arm ({Fukui et al. 2017c}). 
In the region, two clouds elongated along the plane with a velocity separation of 10 \kms \ were found in each of the starbursts to be physically linked {by bridge} features in velocity. The results were interpreted that the two starbursts were formed by triggering in cloud-cloud collision. 
Recently, a similar cloud-cloud collision scenario was presented for the other popular \HII \ regions including M20, M8, M17, and M16 in the {Carina-}Sagittarius Arm (Torii et al. 2011, 2017; {Nishimura et al. 2017a, 2017b) and many star forming regions (Hasegawa et al. 1994,{Sato et al. 2000, Okumura et al. 2001, Miyawaki et al. 1986, 2009, Shimoikura \& Dobashi 2011, Shimoikura et al. 2013, Nakamura et al. 2012, 2014, Dobashi et al. 2015}, Fukui et al. 2015, Torii et al. 2015, Tsuboi et al. 2015, Baug et al. 2016, Dewangan et al. 2017, Gong et al. 2017, Saigo et al. 2017, Fukui et al. 2017b, Fukui et al. 2017e, Kohno et al. 2017, Torii et al. 2017a, 2017b,{2017c}, Hayashi et al. 2017, Sano et al. 2017a, Sano et al. 2017b, Tsutsumi et al. 2017, Ohama et al. 2017a, Ohama et al. 2017b, Ohama et al. 2017c)}. {With the Galaxy scale numerical simulations, cloud-cloud collisions are suggested to be frequent events of every $\sim$ 10 Myr in the inner spiral arm due to the Galactic shear motion and high cloud density (Tasker \& Tan 2009, Dobbs et al. 2015) and the Carina--Sagittarius Arm is the best place to observe these events because it is the nearest inner spiral arm from the Sun.}
It is therefore an issue of keen interest if there are more regions where O star formation is triggered by cloud-cloud collision in the rest of the {Carina-}Sagittarius Arm.

In the present study, we aimed to reveal a large-scale distribution of the molecular gas with a $90''$ beam in the region of GM 24 and mapped an area of $35'\times40'$, which is $100$ times larger than the previous mapping with SEST. The paper is organized as follows; Section 2 describes details of observations {with NANTEN2} and Section 3 the CO results as well as a comparison with the cluster. Section 4 gives discussion on a possible scenario of cloud-cloud collision which triggered formation of the cluster, and Section 5 concludes the paper.

%% file: s2_observations.tex
\section{Observations}
We made \COa\ \Ja \ and \Jb \ observations with the NANTEN2 4 m millimeter/sub-millimeter telescope of Nagoya University.
{Observations} of \COa\ \Ja \ emission were conducted from May 2012 to December 2012. 
The front end was a 4 K cooled SIS mixer receiver.
The system temperature including the atmosphere was $\sim 120$ K in the double-side band (DSB) toward the zenith.
The backend was a digital-Fourier transform spectrometer (DFS) with 16384 channels of 1 GHz bandwidth. 
The velocity coverage and resolution were $\sim 2600$ \kms \ and $0.16$ \kms, respectively. 
We used the on-the-fly (OTF) mapping mode. 
The pointing accuracy was confirmed to be better than $10''$ with daily observations toward the Sun.
The absolute intensity calibration was applied by observing IRAS 16293--2422 [$\alpha_{\rm(J2000)} = \timeform{16h32m23.3s}, \delta_{\rm(J2000)} = \timeform{-24D28'39. 2"}$].
The final beam size was $180 \arcsec$ (FWHM).
The typical rms noise level was $\sim$ 1.2 K ch$^{-1}$. 
We also used \COa\ \Ja \ data cube obtained with {NANTEN}, which was published in \citet{miz04} and \citet{tak10}.
The data were taken with a beam size of \timeform{2.6'} (FWHM) at a \timeform{4.0'} grid spacing and the rms noise level was $\sim$ 0.45 K ch$^{-1}$ with a velocity resolution of 0.65 km s$^{-1}$. 

Observations of \COa\ \Jb \ emission were carried out from October 2014 to November 2015. The typical system temperature was $\sim180$ K (in DSB) including the atmosphere toward the zenith. We observed GM 24 with the OTF mapping mode.  The backend was the same DFS as in $J=$ 1--0 observations.
The velocity coverage and resolution were $\sim$ 1300 km s$^{-1}$ and 0.08 km s$^{-1}$, respectively. The pointing accuracy $\sim$ \timeform{10"} was confirmed by observing the Moon. 
The absolute intensity was calibrated by observing M17SW[$\alpha_{\rm J2000} =\timeform{18h20m24.419s}, \delta_{\rm J2000}$ = \timeform{-16D13'17.628"}]. 
The data cube was smoothed with a Gaussian kernel of \timeform{60"}, and the final beam size was \timeform{100"} (FWHM). The typical rms noise level was $\sim$ 1.1 K ch$^{-1}$.

%% file: s3_results_arxiv.tex
\section{Results}

{Figure 1(a) shows a large-scale distribution of the \COa\ \Jb \ integrated intensity map.
The molecular cloud is distributed along the galactic plane.} GM 24 is located toward $l=$ \timeform{350.4D}--\timeform{350.8D} and {NGC 6334} toward $l=$ \timeform{351.0D}-- \timeform{351.7D}. {These star forming regions are connected with the elongate structure of the CO emission.
Figure 1(b) shows the longitude-velocity diagram. }
Obviously, GM 24 and {NGC 6334} are connected by CO emission, part of the {Carina-}Sagittarius Arm, {in a velocity range from $-8$ \kms \ to 0 \kms at $(l,b) \sim$(\timeform{350.9D}, \timeform{0.8D}).} This suggests that GM 24 and {NGC 6334} are physically linked, which is consistent with their similar distances, 1.75 kpc and 2.0 kpc, from the Sun (e.g., Russeil et al. 2012, Torrelles et al. 1983). 
{NGC 6334 and GM 24 has peak intensity at $\sim -5$ \kms and $\sim -10$ \kms, respectively, which are consistent with the previous observations (Kraemer \& Jackson 1999, G\'omez et al. 1990). In GM 24, there are clouds with two velocity components at $-10$ \kms and $-6$ \kms shown in Figure 1(b). 
The peak column densities are $2\times10^{22}$ \cmcm and $6 \times 10^{21}$ \cmcm with the integration velocity ranges of $-14$-- $-6$ \kms and $-6$ -- $0$, respectively, derived from the $^{12}$CO $J=$ 1--0 data using the $X_{\rm CO}$ conversion factor of $1.0 \times 10^{20}$ (K \kms)$^{-1}$ cm$^{-2}$ (Okamoto et al. 2017). We defied the individual cloud boundary by drawing contours at integrated intensities of 15 K km s$^{-1}$. The total molecular masses are derived as $3 \times10^4$ \Msun and $3 \times10^4$ \Msun, respectively. The cluster and cloud parameters are listed in Table 1.}

{Figure 2(a)} shows an integrated intensity distribution of the \COa \ \Jb \ emission toward GM 24 over a velocity range from $-14$ \kms \ to $-6$ \kms.
{The cloud has a size of 10 pc $\times$ 15 pc {with} two peaks, “a” at $(l,b) \sim$ (\timeform{350.5D}, \timeform{0.96D}), and “b” at $(l,b) \sim$ (\timeform{350.75D}, \timeform{0.95D}).} 
We also see a marked intensity depression toward $(l,b) \sim$ ({\timeform{350.62D}}, \timeform{0.84D}).
{Figure 2(b) shows comparison between $^{12}$CO $J=$2--1 emission and the infrared three color image of {\it Herschel} Space Observatory (Molinari et al. 2010), where blue, green, and red correspond to the 70 $\mu$m, 250 $\mu$m, and 500 $\mu$m emissions, respectively. {The distribution of molecular cloud shows morphological correspondence with the infrared emissions. The CO intensity peaks coincide with those of the infrared emissions, namely that the cluster center is the brightest in the infrared and in CO. On the other hand, the contours of the extended CO emission delineate the faint infrared emission in GM 24.}}

Figure 3 shows velocity channel distributions {of} every 1.2 \kms. {The two peaks of “a” and “b” are detected in {rather large} velocity range from $-15$ km s$^{-1}$ to $-6.6$ \kms , while the molecular cloud distribution exhibits a cavity-like structure at $(l,b) \sim$ (\timeform{350.5D}, \timeform{0.9D}) from $-6.6$ km s$^{-1}$ to $-3.0$ \kms.}
Relations of these features are discussed in Section 4.

%% file: s4_discussion_arxiv.tex
\section{Discussion}
\subsection{A cloud-cloud collision scenario}

In the present study we obtained detailed \COa\ \Jb \ distribution of the molecular gas toward a young high-mass cluster GM 24.
GM 24 is characterized by its compact distribution {with the stellar density of 1000 pc$^{-3}$} similar to the super star clusters which are presumably formed by triggering in cloud-cloud collision (Westerlund 2, \cite{fur09, oha10}; NGC3603, \cite{fuk14}; RCW38, \cite{fuk16}; R136, \cite{fuk17a}; [DBS2003]179, Kuwahara et al. in preparation){, although the number of O stars is small. We summarized the physical parameters of GM 24 and other super star clusters in Table 2.}
It was also suggested that the two starbursts {NGC 6334} and {NGC 6357} next to GM 24 were formed by triggering under cloud-cloud collision ({Fukui et al. 2017c}). {We suggest that the formation of high stellar density in GM 24 requires external compression, while it is unlikely to be formed by the feedback from other high-mass stars because it is isolated. Therefore, we suggest triggering by a cloud-cloud collision as the origin of the compact and dense cluster.}
Considering these scenarios, we test a possibility of cloud-cloud collision in GM 24 as a trigger of the cluster formation in the following. 

\citet{fuk17b} summarized and discussed possible observational signatures of cloud-cloud collision; these authors argued that  two typical  signatures of cloud-cloud collision are a bridge feature and a complementary spatial distribution between the two velocity components {in collision}.
It was also noted that the cloud formed by cloud-cloud collision does not always show double-peaked line profiles, because the two colliding clouds often merge into a single peak, losing its dual {kinematical character}. 
This usually happens for two clouds having different gas column density, and the higher column density cloud becomes dominant in the merged cloud. 
The projection effect, which causes the {observed velocity separation smaller, also tends to apparently make} the two peaks of the colliding clouds into one.

In Figure 3 following the guideline of finding complementary distribution at two velocities (Fukui et al. 2017b), we examined the CO distribution and found that the GM 24 cloud consists of two velocity components with significantly different spatial distributions at $\sim$ $-10$ \kms and $\sim$ $-6$ \kms. According to an eye inspection, the two components show complementary distribution with a displacement in space as shown in Figure 4. {We note that the extended emission at velocity larger than $-6$ km s$^{-1}$ shows no sign related to the present collision, and do not consider them in the following.} Figure 4a shows an overlay of the two components and Figure 4b shows the two components where a displacement is applied, which aims to correct for the collisional motion tilted to the line of sight (Fukui et al. 2017b); the displacement is estimated to be 3 pc {along the galactic plane by applying the method of maximizing an overlapping integral in a complementary distribution between the small cloud and the cavity produced in the large cloud \citep{fuk17b}. In Figure 4b, the two components after the displacement match well with each other; the blue-shifted component coincides with the red-shifted component including the intensity depression at $l=$ \timeform{350.6D} and other features.} {We present more details in Section 4.2 on a comparison between a model of cloud-cloud collision and observations.}

\subsection{Simulated properties of {two} colliding clouds}
{In cloud-cloud collision it might be expected to see two distinct clouds of a narrow linewidth without mixing, whereas the actual distribution in the simulations shows merged clouds which fill the velocity gap between them. We will therefore find a single-peaked relatively broad spectrum whose velocity span is given by a projected velocity separation of the two clouds instead of two discrete colliding clouds as detailed below. }

In order to provide an insight on observational properties of the collision, we describe the physical states of colliding clouds based on the numerical simulations by \citet{tak14}.
The simulations deal with head-on collision between a small cloud and a large cloud, which are spherically symmetric.
We adopt a model listed in {Table 3} for discussion; the radius of the small cloud is 3.5 pc and that of the large cloud 7.2 pc.
The two are colliding at 7 \kms currently at 1.6 Myr after the onset of collision and have internal turbulence in the order of 1--2 \kms \ with highly inhomogeneous density distribution. 
For more details see Takahira et al. (2014, 2017). 
The cloud parameters do not correspond exactly to the present cloud with a factor of 2 difference is size and relative velocity although we do not consider the difference affects a qualitative comparison below.

Figure 5 shows a schematic view of the collision seen from the perpendicular direction to the cloud relative motion. 
In the plane of Figure 5, $\theta$ is an angle of the line of sight to the relative motion of the clouds. We assumed $\theta=$ \timeform{45D} and an epoch of 1.6 Myr after the onset of the collision {considering the projection effect between observed line of sight velocity separation and the actual relative velocity by the random cloud motion}. The epoch is representative, showing cloud signatures typical to collision. The small cloud is producing a cavity in the large cloud by the collisional interaction. The interface layer of the two clouds has enhanced density by collision, where the initial, as well as the collision-induced, turbulence mixes the gas distribution. The layer is on average decelerated to a velocity around 4 \kms as compared with the initial velocity difference by momentum conservation, and the gas in the two clouds is continuously merging into the layer during the collision. {The two clouds are divided into three sections A, B and C in observations as shown in Figure 5; A has the cavity in the large cloud, B the small cloud and the large cloud with the cavity, and C the large cloud alone.}

Figure 6 shows the velocity channel distributions every 1 \kms seen from \timeform{45D} in Figure 5. The small cloud is seen at a velocity range from $-4.1$ to $-2.2$ \kms, and the large cloud from $-1.2$ to $1.7$ \kms. The velocity range from $-2.2$ to $-1.2$ \kms corresponds to the intermediate velocity layer which was created by merging. Note that the velocity ranges of each panel in Figure 6 do not exactly fit the velocity ranges of the two clouds {due to turbulent mixing.} The small cloud is flattened perpendicular to the traveling direction due to the merging into the intermediate layer, and the large cloud has intensity depression corresponding to the cavity created by the small cloud.

Figure 7a shows a position-velocity diagram taken in the direction of the relative motion of the clouds. The clouds as a whole show a{ “V-shape”} as indicated by the thin dashed lines. The main peak is found in $X =$ 5 -- 7 pc and $V/V_0 = -0.6$ -- $-0.2$ \kms. This is a combination of the small cloud and the intermediate layer which merged together. There is an intensity depression in $X = 4$ -- 6 pc and $V/V_0 = -0.3$ -- 0.0 \kms. This corresponds to the cavity in the large cloud. The correspondence between these features in Figures 6 and 7a seems to be reasonable qualitatively by considering that Figure 7a is an average in Y over 0.8 pc.

\subsection{Comparison with the CO data}
{The GM 24 cloud is more complicated than the model and has two peaks {“a”} at $(l, b) =$ (\timeform{350.5D}, \timeform{0.96D}) and  {“b”} at (\timeform{350.75D}, \timeform{0.95D}) (Figure 2(a)). Figure 7b shows the present longitude-velocity diagram in GM 24. In Figure 7b, the primary peak at $l =$ \timeform{350.5D} and $V = -10$ \kms is the small cloud which created the cavity in the large cloud at $l =$ \timeform{350.6D} and $V = -6$ \kms with a projected velocity difference of $\sim 4$ \kms and a displacement of $ 3$ pc (\timeform{0.1D} in galactic longitude). Another peak is seen at $l =$ \timeform{350.75D} and $V = -8$ \kms. We suggest that this peak created a secondary cavity at $l =$ \timeform{350.82D} and $V = -4$ \kms in collision. We also note that the two peaks in the red-shifted cloud {“c”} at $(l,b) =$ (\timeform{350.65D}, \timeform{1.05D}) and {“d”} at (\timeform{350.75D}, \timeform{0.9D}) correspond to the intensity depressions of the blue-shifted cloud, respectively, after the displacement (Figure 4b). {By taking into account the additional peaks observed, {“b”, “c” and “d”},} the present model offers a consistent explanation of the double-peaked distribution of the blue-shifted cloud and the double-depressed distribution of the red-shifted cloud by a single displacement.
}

The complementary distribution found in GM 24 is one of the clearest cases among the known cloud-cloud collisions including M43 \citep{fuk17b}, M17 ({Nishimura et al. 2017a}), and R136 \citep{fuk17a}. 
This clearness is owing to that cloud dispersal by ionization is not significant because of the lower ultraviolet photon flux of the O9 star combined with high visual obscuration of 50--90 mag in the peak of the GM 24 molecular cloud. 
{The GM 24 cloud is also relatively simple with little contamination, and may provide a template in searching for cloud-cloud collision in other regions. }

{The ratio between the displacement and the velocity, 4.2 pc/6 \kms, gives a collision timescale of $\sim$ 0.7 Myr, where the relative motion is assumed to be \timeform{45D} to the line of sight. For a typical mass accretion rate in the collision compressed layer $\sim 10^{-4}$\Msunyr (Inoue and Fukui 2013, {Inoue et al. 2017}), a star with 20\Msun is formed in $2 \times 10^5$ yr well within the time scale. We summarize the physical parameters of the GM 24 cloud; the two cloud have masses ({$3 \times10^4$ \Msun, $3 \times10^4$ \Msun}) and the peak column densities ($2\times10^{22}$ \cmcm, $6 \times 10^{21}$ \cmcm), for the blue-shifted cloud and the red-shifted cloud, respectively, where the $^{12}$
 CO $J=$ 1--0 data were used with an $X_{\rm CO}$ factor of $1.0 \times 10^{20}$ (K \kms)$^{-1}$ cm$^{-2}$ (Okamoto et al. 2017). The high column density $2\times10^{22}$ \cmcm \ is between the initial condition for more than ten O-star formation in the super star clusters like RCW38 \citep{fuk16} and that for a single O star formation like in M20 \citep{tor11, tor17}, and is consistent with the formation of a single O star along with the $\sim100$ low mass stars. 
{The possibility of shock triggered formation of GM 24 from other H\,\emissiontype{II} regions and supernova remnants seems to be unlikely because GM 24 is an isolated cluster. Therefore, we suggest that the formation of the compact cluster with high stellar density requires a local compression with an external triggering process like a cloud-cloud collision.  }
The compactness of the GM 24 cluster less than 0.3 pc is notable, suggesting a strong compression which operated to form the cluster. 
Given the molecular mass of $\sim200$ \Msun \ within 0.3 pc of the cluster the star formation efficiency is estimated to be as high as 50 percent. 
This suggests that the collisional impact is powerful in compressing the molecular gas even if a collision velocity is not highly supersonic at $\sim 6$ \kms.
{ In the previous paper, cloud-cloud collision was suggested as a trigger of formation of {NGC 6334} and {NGC 6357} ({Fukui et al. 2017c}). The trigger was made between two elongated clouds of 100-pc length in the Carina-Sagittarius Arm. The trigger in GM 24 took place within the red-shifted component of these two elongated clouds at a smaller relative velocity, showing a large variety of cloud-cloud collision.}

%% file: s5_conclusions.tex
\section{Conclusions}
We have carried out new CO $J=2-1$ observations toward GM 24.
The main conclusions are summarized below. 

\begin{enumerate}
\item 
GM 24 is associated with a molecular cloud having $\sim10^4$ \Msun \ as shown by the previous observations \citep{tap85}.
The cloud consists of two velocity components at $-10$ \kms \ and $-6$ \kms.
The $-10$ \kms \ component has a high column density of $2\times10^{22}$ \cmcm \ with a total mass of ${3 \times 10^4}$ \Msun, while the $-6$ \kms \ component has a column density of $6\times10^{21}$ \cmcm \ with a total mass of  ${3 \times10^4}$ \Msun.

\item 
The two components show spatially complementary distribution typical to two colliding clouds. 
A relative displacement of the clouds along the plane by 3 pc provides a reasonable fit between the blue-shifted cloud and the intensity depression of the red-shifted cloud including the secondary peak and depression.
By assuming that the relative motion makes an angle of 45 degrees to the line of sight, the collision timescale is calculated to be $\sim 10^5$ yr.

\item 
The cluster is compact with a size of 0.3 pc around the most massive star Irs3, an O9 star, including 100 stars, and is coincident with the compact CO peak of the blue-shifted cloud with a similar small size.
This suggests that the cluster was formed in the blue-shifted cloud under a strong collisional impact by the red-shifted cloud.
The formation seems qualitatively common to the formation of the compact super star clusters.
Since the cluster includes lower mass members as well, the present case may suggest that low mass stars, not only high mass stars, are also formed by triggering in cloud-cloud collision.

\end{enumerate}

\section*{Acknowledgements}
NANTEN2 is an international collaboration of ten universities: Nagoya University, Osaka Prefecture University, University of Cologne, University of Bonn, Seoul National University, University of Chile, University of New South Wales, Macquarie University, University of Sydney, and Zurich Technical University.

The work is financially supported by a Grant-in-Aid for Scientific Research (KAKENHI, No. 15K17607, 15H05694) from MEXT (the Ministry of Education, Culture, Sports, Science and Technology of Japan) and JSPS (Japan Soxiety for the Promotion of Science).

%% file: s9_figures_arxiv.tex
\begin{figure*}[h]
\begin{center}
\includegraphics[width=14cm]{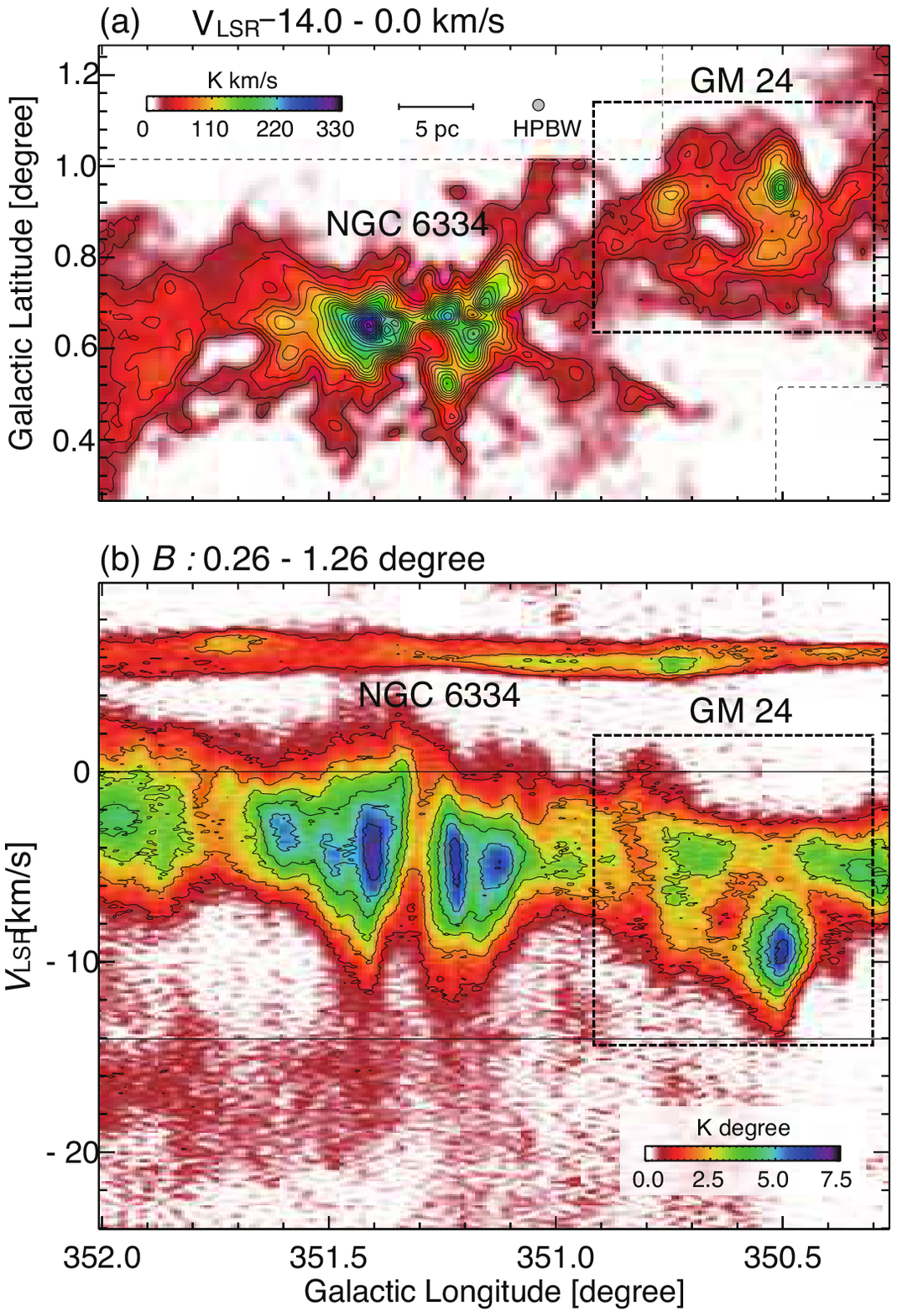}
\end{center}
\caption{(a) Intensity map of $^{12}$CO($J$ = 2--1) obtained with NANTEN2. The integration velocity range is from $-14.0$ to 0.0 km s$^{-1}$. The lowest contour and contour intervals are 30 K km s$^{-1}$ and 15 K km s$^{-1}$, respectively. (b) Galactic Longitude--Velocity diagram of $^{12}$CO($J$ = 2--1). The integration range of Galactic Latitude is from $0.26$ to 1.26 degree. The lowest contour and contour intervals are 1 K degree. {}}
\label{fig1}
\end{figure*}%

\begin{figure*}[h]
\begin{center}
\includegraphics[width=16cm]{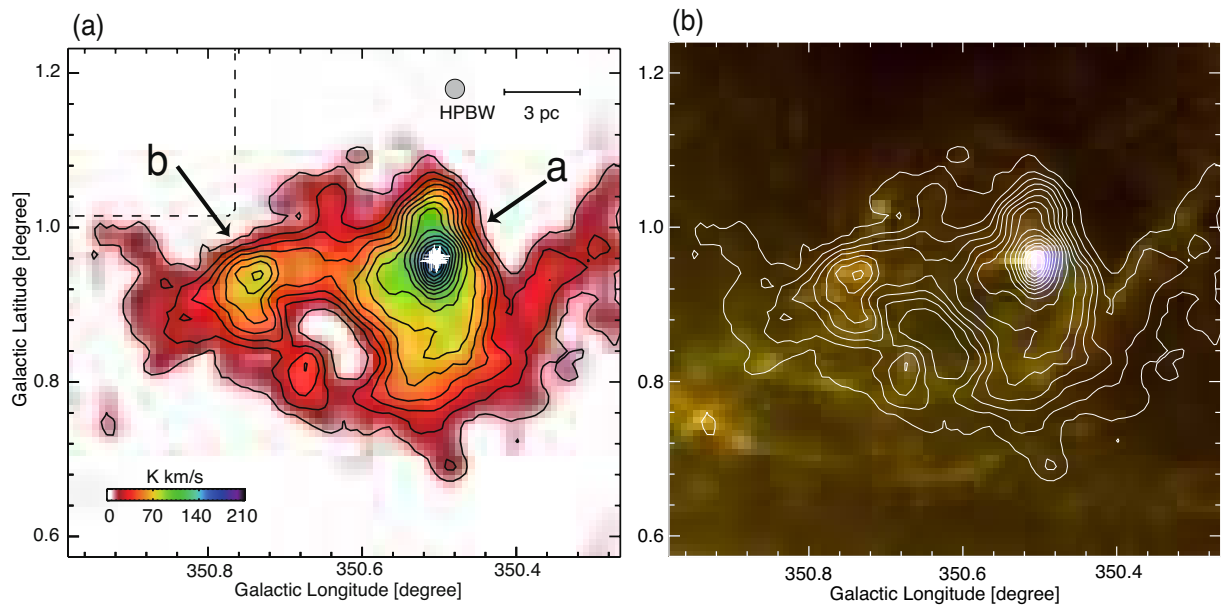}
\end{center}
\caption{(a) Intensity map of $^{12}$CO($J$ = 2--1) toward GM~24. The integration velocity range is from $-14.0$ to $-6.0$ km s$^{-1}$. The lowest contour and contour intervals are 12 K km s$^{-1}$. The open crosses represent to the positions of YSO cataloged by Tapia et al (1991). Two peaks{of “a” and “b” are indicated. (b) Three color infrared image of GM~24 with blue, green, and red corresponding to {\it Herschel}/PACS 70 $\mu$m, {\it Herschel}/SPIRE 250 $\mu$m, and {\it Herschel}/SPIRE 500 $\mu$m, respectively (Molinari et al. 2010). The contour is same of (a).}}
\label{fig2}
\end{figure*}%

\begin{figure*}[h]
\begin{center}
\includegraphics[width=14cm]{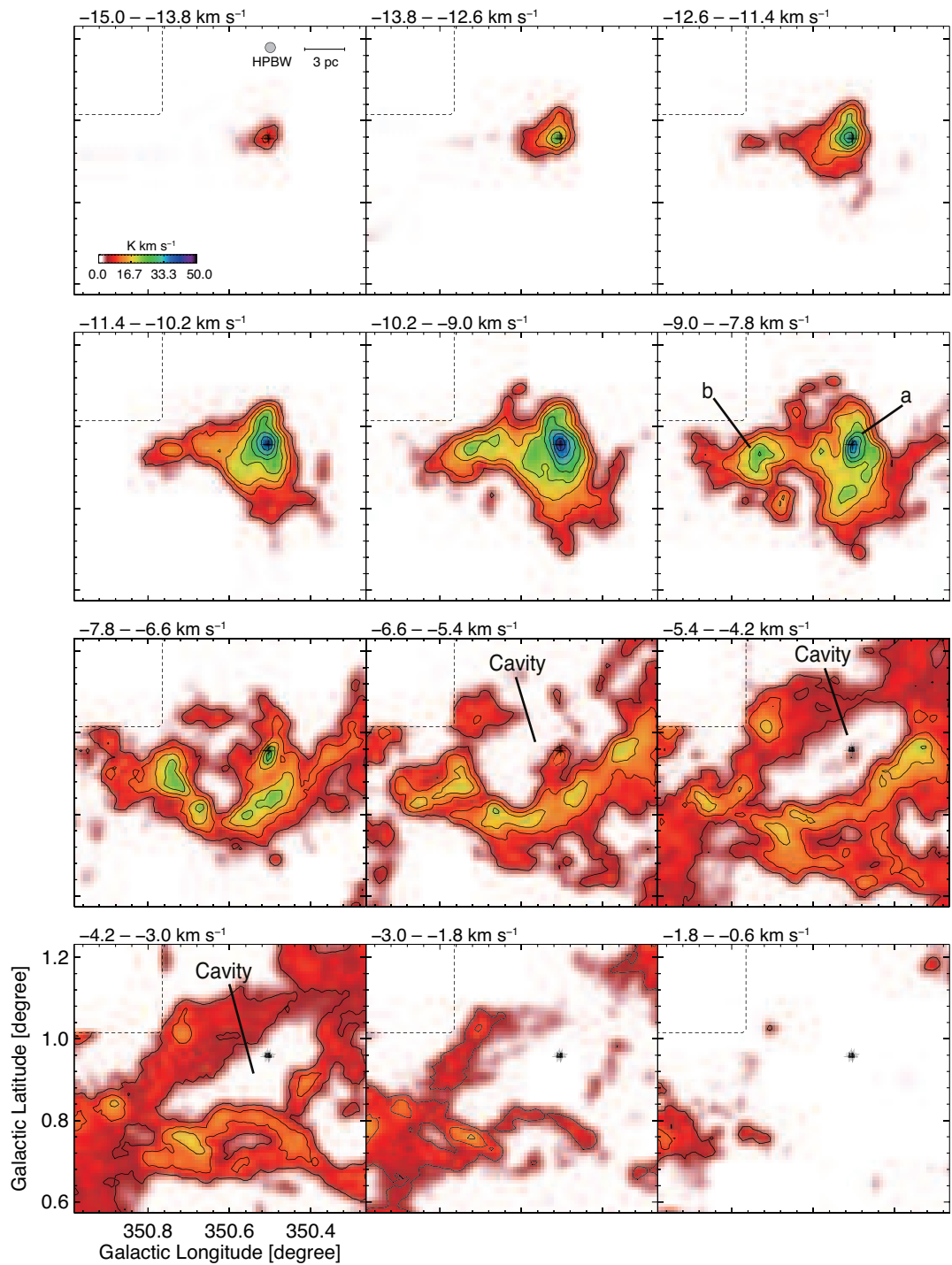}
\end{center}
\caption{Velocity channel maps of $^{12}$CO($J$ = 2--1) toward GM~24. Each panel shows a CO i ntensity integrated over the velocity range from $-36.6$ to $6.6$ km s$^{-1}$ every 1.2 km s$^{-1}$. The lowest contour and contour intervals are 5 K km s$^{-1}$. The open crosses are the same as in {Figure 2(a)}.}
\label{fig3c}
\end{figure*}%

\begin{figure*}[h]
\begin{center}
\includegraphics[width=9cm,clip,angle=90]{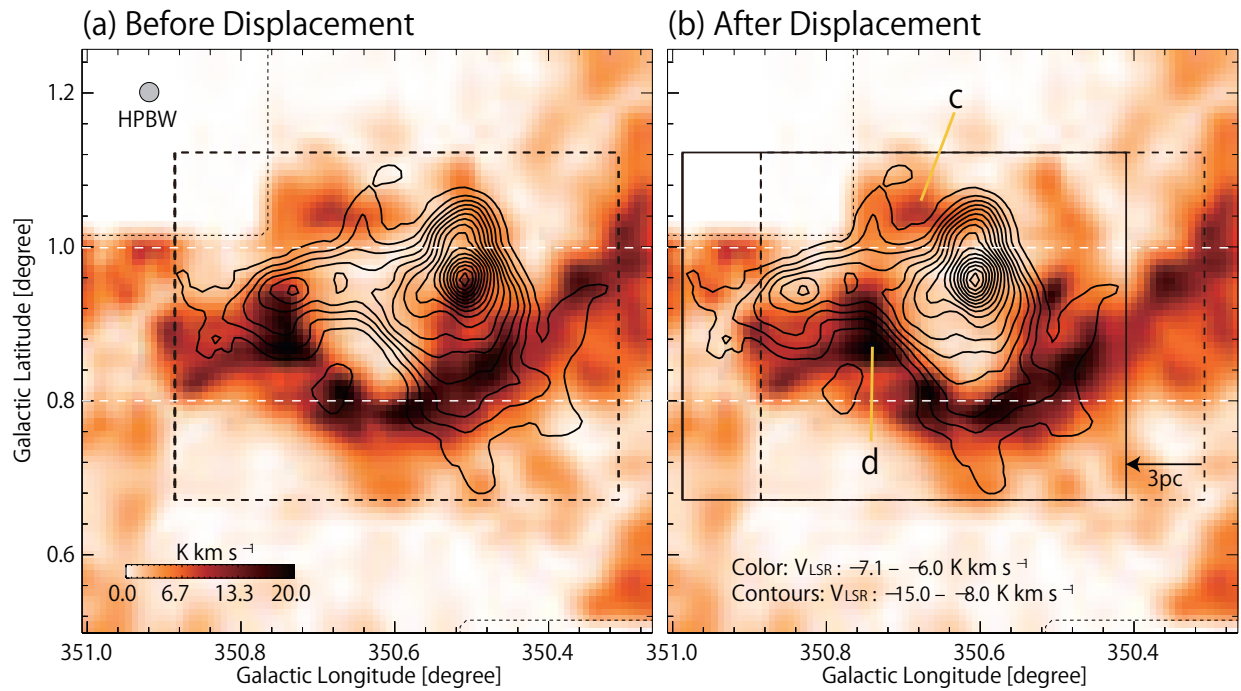}
\end{center}
\caption{(a) Complementary $^{12}$CO($J$ = 2--1) distributions of the two velocity components of GM~24. The integrated velocity range is from $-7.1$ to $-6.0$ km s$^{-1}$ for color image; $-15.0$ to $-8.0$ km s$^{-1}$ for contours. The lowest contour and contour intervals are 10 K km s$^{-1}$. The color image and contours indicate the red- and blue-shifted clouds, respectively. Figures \ref{fig4}(a) and \ref{fig4}(b) correspond to the maps before and after displacement of the blue-shifted cloud contours. Two peaks {“c” and “d”} in the red-shifted cloud are indicated.}
\label{fig5}
\end{figure*}

\clearpage

\begin{figure*}[h]
\begin{center}
\includegraphics[width=7cm]{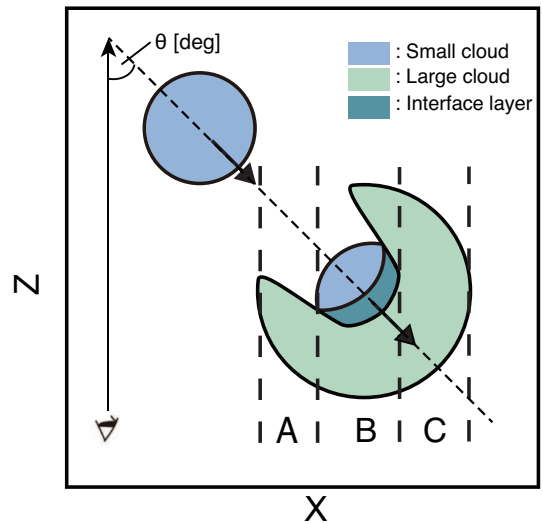}
\end{center}
\caption{Schematic of the top-view of the collision; [A] the cavity, [B] the cavity and the small cloud, and [C] the large cloud.}
\label{fig4}
\end{figure*}

\begin{figure*}[h]
\begin{center}
\includegraphics[width=17cm]{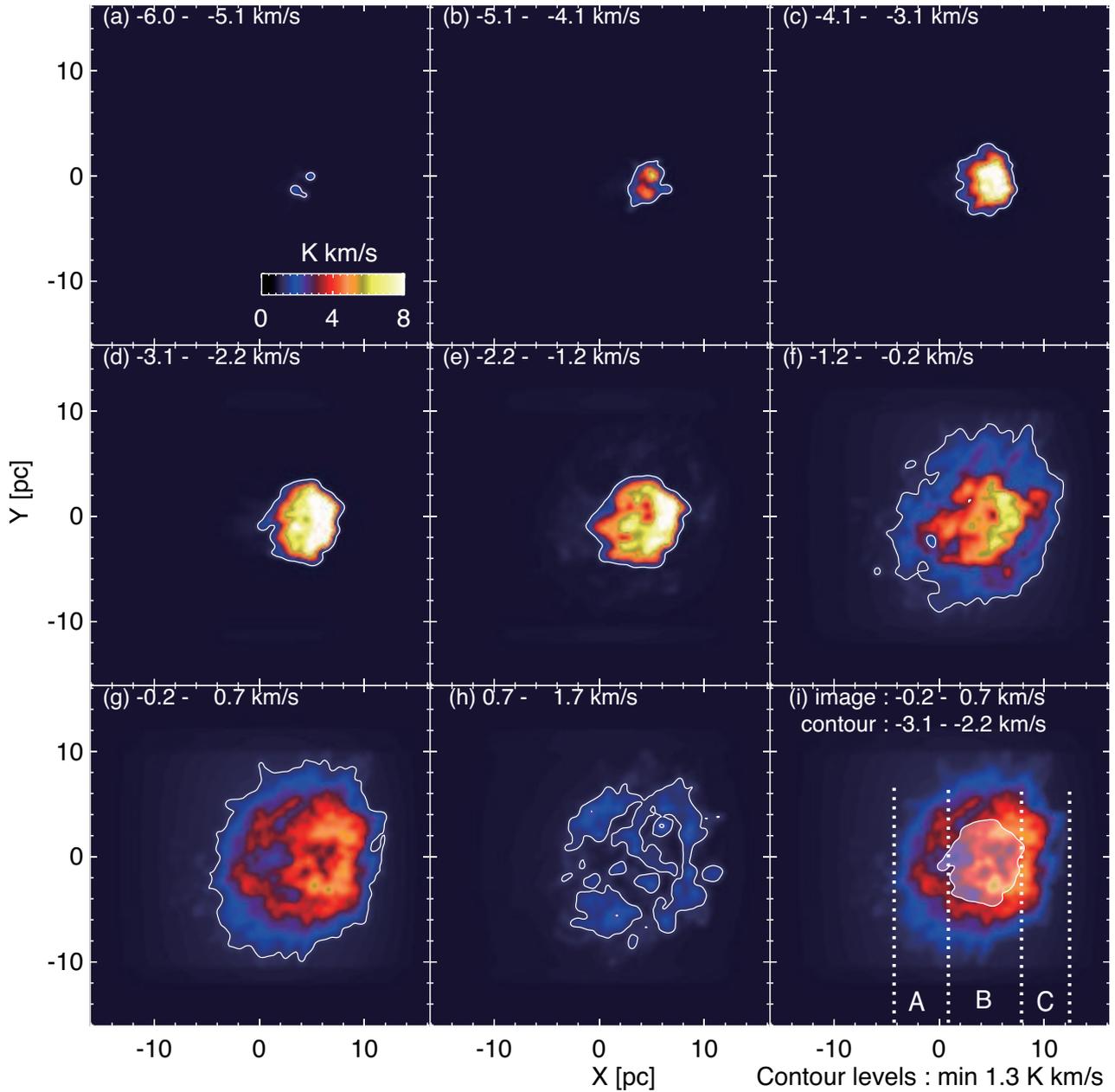}
\end{center}
\caption{Systematic observations of $^{12}$CO $J=$ 1--0 emission based on the numarical simulations by Takahira et al. (2014) observed at an angle of the relative motion to the line of sight $\theta=45\deg$.
(a)-(h) show the velocity channel distributions every 0.96 \kms map simulation model. The parameters of the model are shown in Table 1.(i) shows a complementary distribution between the large cloud, the image in (g), and the small cloud with the contour of (d) at 1.3 K \kms}
\label{fig5}
\end{figure*}

\begin{figure*}[h]
\begin{center}
\includegraphics[width=7cm]{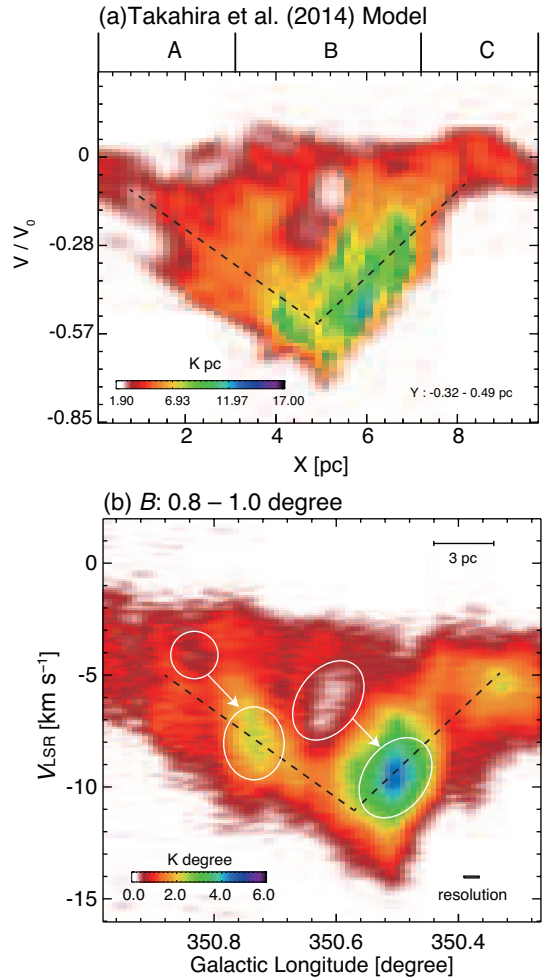}
\end{center}
\caption{(a) Position-velocity diagram of the cloud-cloud collision model. The image were reproduced by using Takahira et al. (2014) model. (b) Galactic Longitude--Velocity diagram of $^{12}$CO($J$ = 2--1) toward GM~24. The integration range of Galactic Latitude is from 0.8 to 1.0 degree. }
\label{fig4}
\end{figure*}

%% file: s9_tables.tex
\begin{table}[h] 
\begin{center}
\tbl{The physical parameters of GM 24}{% 
\begin{tabular}{@{}cccccc@{}} \noalign{\vskip3pt}
\hline\hline
\multicolumn{1}{c}{Properties of the stellar cluster$^{\dag}$} &  &   \\ 
\hline
Size [pc] & 0.39 &    \\
Number of YSOs & $>100$ &  \\
Total stellar mass [$M_{\odot}$] & 450 &  \\
Total luminosity [$L_{\odot}$] & $4 \times 10^4$ &  \\
\hline
Properties of molecular cloud & The $-10$ km s$^{-1}$ cloud & The $-6$ km s$^{-1}$ cloud     \\
\hline
Column densities [cm$^{-2}$] & $2 \times 10^{22}$ & $6 \times 10^{21}$   \\
Mass [$M_{\odot}$] & $3 \times 10^{4}$ & $3 \times 10^{4}$   \\
\hline\noalign{\vskip3pt} 
\end{tabular}} 
\label{tab:first} 
\begin{tabnote}
\footnotemark[$\dag$] References : Tapia \& Persi (2008), Tapia et al. (2009)
\end{tabnote} 
\end{center} 
\end{table}

\begin{table}[h] 
\begin{center}
\tbl{Comparisons between GM 24 and Super Star Clusters}{% 
\begin{tabular}{@{}ccccccc@{}} \noalign{\vskip3pt}
\hline\hline
\multicolumn{1}{c}{Name} & Cloud Masses &  Column Densities &  Cluster Age & \# of O stars & Reference \\ [2pt]
\noalign{\vskip3pt}
 &  [$M_{\odot}$] & [cm$^{-2}$]  & [Myr] &  & \\
(1)  & (2)  & (3)  & (4)  & (5)  & (6) \\
\hline
GM 24 & ($3 \times 10^4, 3 \times 10^4$) &  ($2 \times 10^{22}, 6 \times 10^{21}$) & $<$1 \footnotemark[$*$] & $> $1 & This study \\
RCW38 & ($2 \times 10^4, 3 \times 10^3$) &  ($1 \times 10^{23}, 1 \times 10^{22}$) & $\sim 0.1$ & $\sim 20$ &[1] \\
NGC3603 & ($7 \times 10^4, 1 \times 10^4$)  & ($1 \times 10^{23}, 1 \times 10^{22}$) & $\sim 2$ & $\sim 30$ & [2]\\
Westerlund2 & ($9 \times 10^4, 8 \times 10^4$)  & ($2 \times 10^{23}, 2 \times 10^{22}$) & $\sim 2$ & $14$ &  [3,4]\\
$[$DBS2003$]$179  & ($2 \times 10^5, 2 \times 10^5$)  & ($8 \times 10^{22}, 5 \times 10^{22}$) & $\sim 5$ &  $>10$ & [5] \\
\hline\noalign{\vskip3pt} 
\end{tabular}} 
\label{tab:first} 
\begin{tabnote}
\footnotemark[] Note. Columns: (1) Name. (2,3) Molecular masses and column densities of the two colliding clouds. (4, 5) Age and number of O stars. (6) References: [1] Fukui et al. (2016) [2] Fukui et al. (2014) [3] Furukawa et al. (2009) [4] Ohama et al. (2010) [5] Kuwahara et al in preparation. \\
\footnotemark[$*$] References : Tapia et al. (2009)
\end{tabnote} 
\end{center} 
\end{table}

\begin{table}[h] 
\begin{center}
\tbl{The parameters of the numerical simulations (Takahira et al. 2014) }{% 
\begin{tabular}{@{}cccccc@{}} \noalign{\vskip3pt}
\hline\hline
\multicolumn{1}{c}{Box size [pc]} & $30 \times 30 \times 30$  &  &  &  \\ [2pt]
\noalign{\vskip3pt}
Resolution [pc] & 0.06 &  &  \\
Collsion velocity $V_0$ [km s$^{-1}$] & {10 (7)$^{\dag}$} & & \\
\hline
Parameter & The small cloud & The large cloud & note    \\
\hline
Temperature [K] & 120 & 240 &  \\
Free-fall time [Myr] & 5.31 & 7.29 &  \\
Radius [pc] & 3.5 & 7.2 &  \\
Mass [$M_{\odot}$] & 417 & 1635 &  \\
Velocity dispersion [km s$^{-1}$] & 1.25 & 1.71 &  \\
Density [cm $^{-3}$] & 47.4 & 25.3 & Assumed a Bonner-Ebert sphere  \\
\hline\noalign{\vskip3pt} 
\end{tabular}} 
\label{tab:first} 
\begin{tabnote}
\footnotemark[$\dag$] The initial relative velocity between the two clouds is set to 10 \kms, whereas the collisional interaction decelerates the relative velocity to about 7 \kms in 1.6 Myrs after the onset of the collision. The present synthetic observations correspond to for a relative velocity $V_0=7$ \kms at 1.6 Myrs.
\end{tabnote} 
\end{center} 
\end{table}